\documentclass[12pt,a4paper]{panl}
\usepackage{cite}
\usepackage{wrapfig}
\usepackage{graphicx}
\usepackage{amssymb}
\usepackage{amsfonts}
\usepackage{amsmath}
\usepackage{longtable}
\usepackage{rotating}
\usepackage{lscape}
\usepackage{epsfig}
\usepackage{multirow}

\originalTeX
\begin{document}

\title{Nonperturbative Kinetic Description of the Vacuum QGP Production in the Abelian Projection Method\\
}
\maketitle
\authors{V.V.\,Dmitriev$^{a}$\footnote{E-mail: dmitrievvv@sgu.ru},
S.V.\,Churochkina$^{a}$, V.A.\,Tseryupa$^{a}$, S.A.\,Smolyansky$^{a}$\footnote{E-mail: smol@sgu.ru}}
\setcounter{footnote}{0}
\from{$^{a}$\,Saratov State University, Saratov, Russia}

\begin{abstract}
%
Within the framework of the Abelian projection method in the $SU(N)$ variant of QCD, a system of nonperturbative selfconsistent equations for the back-reaction problem is derived. This system describes the vacuum production of quarks under the influence of strong quasiclassical gluon fields taking into account violating of the electroneutrality condition.

\end{abstract}
\vspace*{6pt}


\section{Introduction}\label{sec:Intro}

In the initial works (\cite{1} and the review \cite{2}), a nonperturbative kinetic description of parton vacuum production in ultrarelativistic heavy-ion collisions (UHICs) utilized an Abelian variant of kinetic theory. This model involved fermions with light quark parameters interacting with a strong Abelian field that simulated the gluon field and was adjusted ad hoc to provide sufficient multiplicity in the out-state.

In the present communication, we return to the problem of the kinetic description of quark vacuum creation and the pre-equilibrium evolution of the quark-gluon plasma (QGP) at the initial stage of UHICs. As a first step, we will limit ourselves to a model based on the correct formulation of the Abelian projection method (APM) in QCD \cite{3}. This is a necessary stage for the subsequent generalization of the kinetic theory towards accounting for non-Abelian degrees of freedom. Another new element is the substitution of the external field concept by self-consistent internal (plasma) gluon fields generated by strong quark currents at the very early stage of deconfinement.

Below, the simplest model situation is considered within the framework of the APM  QCD  with the following additional limitations:
\begin{enumerate}
\item the electrodynamical processes in the evolution of QGP are ignored,
\item the gluon fields are considered in the mean field approximation,
\item it is assumed that the quasiclassical gluon fields are spatially homogeneous (which leads to spatial homogeneity of the quark subsystem) and time dependent,
\item polarizations of all gluon fields are selected identical and fixed.
\end{enumerate}

These limitations are introduced by analogy with strong-field QED (SFQED) \cite{4,5,6} and graphene \cite{7,8}. The last limitation can be canceled, leading to a significant increase in the number of kinetic equations due to the account of the spin degrees of freedom (e.g., \cite{5,9}).

Summing the constraints 1)-4), one obtains the following structure
of the quasiclassical gluon fields in the Hamilton gauge:

\begin{equation}\label{1}
B^C_{\mu} (x) = (0,0,0, B^C_3 (t)=B^C(t)),
\end{equation}
where $C=1,...,N_g=N_c^2-1$ is the gluon color index.

The system of kinetic equations (KEs) for the quark sector of QCD obtained under these conditions is introduced in Sec.~\ref{sec:QKE}, based on a profound structural analogy with Abelian gauge theories (SFQED \cite{4,5,6} and graphene \cite{7,8}). The back reaction problem is formulated in Sec.~\ref{sec:BR}. The obtained results and some open problems are briefly discussed in Sec.~\ref{sec:Sum}.

\section{Quark Kinetic Equation in APM}\label{sec:QKE}

The Dirac equation for the quark subsystem in strong quasiclassical gluon fields $B^{C}_{\mu}$ in $SU(N)$ QCD, presented in a form convenient for discussion, is
\begin{equation}\label{2}
\left[
\gamma^\mu
\left( i \partial_\mu \delta_{ab} + g_{s} B^{C}_\mu(x) t^C_{ab} \right)
m_{q} \delta_{ab}
\right]
\Psi_{q, b}(x) = 0,
\end{equation}
where indices $q=1, ..., N=N_f$ and $a, b=1,..., N=N_c$ correspond to the flavour and color of the quark fields, respectively, while index $C=1,...,N_c^2-1$ denotes the color of the gluon fields; matrices $t_{ab}^C$ are defined via the Gell-Mann matrices $\lambda_{ab}^C = 2t_{ab}^C$; and $g_s$ is the QCD coupling constant.

The APM in $SU(N)$ gauge theory \cite{3,10,11,12,13} emphasizes the special role of Abelian-type gluon fields $B^c_{\mu}(x)$, defined by the subset (Cartan) of diagonal generators of the $SU(N)$ group and are enumerated by the index $c$,
\begin{equation}\label{3}
t^c_{ab} = t^c_a \delta_{ab} .
\end{equation}
The number of diagonal generators is  $N_d = N-1$ \cite{10}. Relation (\ref{3}) ensures color independence of the quark fields during interaction with gluon fields from the class $\{B^c_{\mu}(x)\}$.

Examples of diagonal generators: for $SU(2)$ — the Pauli matrix $\sigma^3$ ($N_d=1$, Abelian-type gluon field $B^{c=3}_{\mu}$); for $SU(3)$ — the Gell-Mann matrices $\lambda^3$ and $\lambda^8$ ($N_d=2$, Abelian-type gluon fields $B^{c=3,8}_{\mu}$).

For the subset of Abelian-type gluon fields, the coupled system (\ref{2}) decouples into a system of $N^2$ independent equations (without summation over $a$):
\begin{equation}\label{4}
\left[
\gamma^\mu
\left( \partial_\mu + i g_s B_{\mu,a}(x) \right)
m_q
\right]
\Psi_{qa}(x) = 0,
\end{equation}
where the effective Abelian-type field is defined as
\begin{equation}\label{5}
B_{\mu,a}(x) = \sum_{c}^{N_d}
B^c_{\mu}(x) t^c_{a}
, \quad a = 1, ..., N_c.
\end{equation}

This decoupling of the original system (\ref{2}) into independent equations (\ref{4}) relies on a specific ad hoc selection of gluon fields and fixes their gauge, but does not introduce $U(1)$ symmetry into Eqs. (\ref{4}). The similarity of Eqs. (\ref{4}) and the Dirac equation in a quasiclassical electromagnetic field is formal. Nevertheless, this mathematical analogy permits the adaptation of corresponding results from strong-field QED (SFQED) under conditions where Eq. (\ref{4}) holds.

Such an adaptation also enables us to use results from nonperturbative kinetic theory describing vacuum production of charged particles under strong time-dependent electric fields in SFQED \cite{4,5,6} and graphene \cite{7,8}. Using this analogy, the system of KEs for the quark subsystem with Abelian-type gluon fields with fixed polarization $\mu=3$ can be written in integro-differential form ($\xi=q,a$):
\begin{equation}\label{6}
  \dot{f}_{\xi}({\bf p},t)
  = \frac{1}{2}
  \lambda_{\xi}({\bf p},t)
  \int_{t_0}^t dt' \
  \lambda_\xi({\bf p},t')
  \left[ 1 - f_\xi({\bf p},t') - \tilde{f}_\xi(-{\bf p},t') \right]
  \cos\theta_\xi({\bf p}; t,t')
\end{equation}
or as a system of ordinary differential equations:
\begin{equation}\label{7}
  \dot{f}_{\xi}({\bf p},t) = \frac{1}{2} \lambda_\xi({\bf p},t) u_\xi({\bf p},t),
\end{equation}
\begin{equation}\label{8}
\dot{u}_{\xi}({\bf p},t) = \lambda_\xi({\bf p},t) \left[ 1 - f_\xi({\bf p},t) - \tilde{f}_\xi(-{\bf p},t) \right] - 2 \varepsilon_\xi({\bf p},t) v_\xi({\bf p},t),
\end{equation}
\begin{equation}\label{9}
\dot{v}_{\xi}({\bf p},t) = 2\varepsilon_\xi({\bf p},t) u_\xi({\bf p},t),
\end{equation}
where $f_\xi({\bf p},t)$ and $\tilde{f}_\xi(-{\bf p},t)$ are the distribution functions of the quarks and antiquarks of the sort $\xi$, $v_\xi({\bf p},t)$ and $u_\xi({\bf p},t)$ are the vacuum polarization functions of the energy and current, respectively,
\begin{equation}\label{10}
  \lambda_\xi({\bf p},t) = g_s \varepsilon_{\perp\xi}({\bf p}) \dot{B}_\xi(t)
  / \varepsilon^2_\xi({\bf p},t)
\end{equation}
is the vacuum excitation amplitude;
\begin{equation}\label{11}
\varepsilon_\xi({\bf p},t)
= \left[ \varepsilon^2_{\perp\xi}({\bf p}) + P^2_{\xi}(t) \right]^{1/2},
\end{equation}
\begin{equation}\label{12}
\varepsilon_{\perp\xi}({\bf p})
= \left[ m^2_q + p^2_{\perp} \right]^{1/2}
\end{equation}
are the quasienergy and transversal energy, $p_\perp$ is the transversal momentum, $P_\xi(t) = p^3 - g_s B_{a}(t)$ is the third component of the quasimomentum with the field (\ref{5}) ($\mu=3$); and finally,
\begin{equation}\label{13}
\vartheta_\xi({\bf p}; t,t') = 2 \int_{t'}^t d\tau \ \varepsilon_{\xi}({\bf p}, \tau)
\end{equation}
is the high-frequency phase.

Derivation details for KEs (\ref{6})-(\ref{9}) can be found in \cite{5,14,15} (SFQED) and \cite{7,8} (graphene).

Typically, kinetic theory based on QED incorporates an electroneutrality condition, ensuring equality between electron and positron distribution functions and leading to the Pauli blocking factor $1-2f({\bf p},t)$. In the present kinetic theory formulation, no such condition exists, and $f_\xi({\bf p},t) \neq \tilde{f}_\xi(-{\bf p},t)$. However, according to the derivation of KEs (\ref{7})-(\ref{9}), quark and antiquark production rates coincide, i.e.,
\begin{equation}\label{14}
\dot{f}_\xi({\bf p},t) = \dot{\tilde{f}}_\xi(-{\bf p},t),
\end{equation}
from which it follows that
\begin{equation}\label{15}
\tilde{f}_\xi(-{\bf p},t) = f_\xi({\bf p},t) + \varphi_\xi({\bf p}) \leqslant 1
\end{equation}
at any time.

Although electrodynamic processes are not included in the present formalism, charge conservation must still be maintained. According to Eq. (\ref{15}), functions $\varphi_\xi({\bf p})$ describe violations of electroneutrality for each quark type at the initial time corresponding to deconfinement. These violations persist also for $t \geqslant t_0$.

The influence of non-compensated charges on vacuum quark production can be incorporated into KEs (\ref{6}) and (\ref{7})-(\ref{9}) as follows:
\begin{equation}\label{16}
  \dot{f}_\xi(t)
  = S_{\xi}(t)
  + \frac{1}{2}
  \lambda_{\xi}(t)
  \int_{t_0}^t dt' \
  \lambda_\xi(t')
  \left[ 1 - 2f_\xi(t') \right]
  \cos\theta_\xi(t,t')
\end{equation}
and
\begin{equation}\label{17}
  \dot{f}_\xi(t) = S_{\xi}(t) + \frac{1}{2} \lambda_{\xi}(t) u_{\xi}(t),
\end{equation}
\begin{equation}\label{18}
  \dot{u}_\xi(t) = \lambda_{\xi}(t) \left[ 1 - 2f_{\xi}(t) \right] - 2\varepsilon_{\xi}(t) v_{\xi}(t),
\end{equation}
\begin{equation}\label{19}
  \dot{v}_{\xi}(t) = 2\varepsilon_{\xi}(t) u_{\xi}(t),
\end{equation}
where
\begin{equation}\label{20}
  S_{\xi}(t) = -\frac{1}{2} \varphi_{\xi} \lambda_{\xi}(t)
  \int_{t_0}^t dt' \
  \lambda_{\xi}(t') \cos\theta_{\xi}(t,t')
\end{equation}
is the source term for vacuum quark production due to non-compensated electric charges in different quark subsystems. For simplicity, momentum arguments are omitted in Eqs. (\ref{16})-(\ref{20}).
\section{The Back Reaction Problem}\label{sec:BR}

Under specific conditions, the back reaction mechanism can play an important role in the pre-equilibrium evolution of the QGP generated in UHICs. In the $SU(3)$ QCD the complete system of equations describing the back reaction mechanism consists of the quark KE system (\ref{16}) (or (\ref{17})-(\ref{19})) coupled with the equations for the quasiclassical gluon fields generated by strong counter-propagating primordial quark currents in the projection on the gluon color space (see below (\ref{25})-(\ref{27})):
\begin{equation}\label{21}
D_{\mu} G_{A}^{\mu\nu}(x)
\equiv \partial_{\mu} G_{A}^{\mu\nu}(x) + g_s f_{ABC} B_{\mu B }(x) G_{C}^{\mu\nu}(x)
= j_{A}^{\nu}(x),
\end{equation}
where
\begin{equation}\label{22}
G_{A}^{\mu\nu}(x)
\equiv \partial^{\mu} B_{A}^{\nu}(x) - \partial^{\nu} B_{A}^{\mu}(x) + g_{s} f_{ABC} B_{B}^{\mu}(x) B_{C}^{\nu}(x)
\end{equation}
and $f_{ABC}$ are the structure constants of the $SU(3)$ group.

Let us adapt Eqs. (\ref{21}) to the considered situation. For a spatially homogeneous system in the gauge (\ref{1}), the field tensor (\ref{22}) becomes
\begin{equation}\label{23}
G_{A}^{30}(t) = -G_{A}^{03}(t) = \dot{B}_{A}(t)
\end{equation}
and consequently, according to Eq. (\ref{21}),
\begin{equation}\label{24}
\ddot{B}_{A} = j_{A}^{\mu=3}
\end{equation}
which corresponds to the Abelian variant of the theory \cite{16}.

Equations (\ref{21})-(\ref{24}) are written in the gluon color space. This remark also applies to the quark currents $j_{A}^{\mu}(x)$ in Eqs. (\ref{21}) and (\ref{24}). The connection with the quark current in the proper color space is given by the relation \cite{17}
\begin{equation}\label{25}
    j_{q\  }^{\mu C}(x)
    = g_s \sum_{a,b} t_{ab}^{C} \langle 0_{in} | \bar{\Psi}_{qa}(x) \gamma^{\mu} \Psi_{q b  }(x) | 0_{in} \rangle.
\end{equation}
Within this framework, one can use Eq. (\ref{5}) to obtain the relation between currents in the color spaces of the quark (right-hand side) and gluon (left-hand side) fields:
\begin{equation}\label{26}
    j^{\mu c}(x) = g_s \sum _{q,a} t^{c}_{a} J^{\mu}_{qa}(x),
\end{equation}
where
\begin{equation}\label{27}
    J^{\mu}_{qa}(x) = \langle 0_{in} | \bar{\Psi}_{qa}(x) \gamma^{\mu} \Psi_{qa}(x) | 0_{in} \rangle.
\end{equation}
The third component of the current (\ref{26}) appears on the right-hand side of Eq. (\ref{24}) for $A=c$.

From Eq. (\ref{24}), one can derive the equation for the effective Abelian-type combinations of the gluon fields (\ref{5}):
\begin{equation}\label{28}
    \ddot{B}_{a}(t) =  g_s \sum_{q,b,c} t_{a}^{c} t_{b}^{c} J_{q,b}^{\mu=3}(t).
\end{equation}
In the considered formulation of the problem, Eq. (\ref{28}) supplements the system of kinetic equations (\ref{16}) (or (\ref{17})-(\ref{19})) to the complete system of equations for the back reaction problem. The quark current density (\ref{27}) can be expressed in terms of the basic functions $f_{\xi}({\bf p},t)$ and $u_{\xi}({\bf p},t)$ using kinetic theory \cite{16,18}.

\section{Conclusion}\label{sec:Sum}

In contrast to the initial series of works (reviewed in \cite{2}) on the nonperturbative kinetic description of vacuum parton production in a strong $U(1)$ gauge field, the present theoretical formalism is a correct implementation of the APM for $SU(N)$ QCD, within which a closed system of self-consistent equations accounting for back reaction processes has been derived. This advancement not only increases the number of relevant equations but also introduces some features (for example the violation of the electroneutrality condition). The presented model already provides perspectives for further investigation of the pre-equilibrium stage of QGP evolution (see, e.g., \cite{19}).

\section*{Acknowledgements}The authors are grateful to A.S. Parvan for his attention to and support of this work.

\section*{Conflict of interest}The authors declare no conflict of interest.

\end{document}